\documentclass[preprint]{aastex}

\def\ltsim{ \,{}^<_\sim\, }
\def\gtsim{ \,{}^>_\sim\, }

\shorttitle{Galaxy Halo Formation}
\shortauthors{Harris \& Harris}

\begin{document}

\title{On the Formation of Galaxy Halos: \\
Comparing NGC 5128 and the Local Group Members}

\author{William E.~Harris}
\affil{Department of Physics \& Astronomy, McMaster University,
    Hamilton ON L8S 4M1 }
\email{harris@physics.mcmaster.ca}

\author{Gretchen L.~H.~Harris}
\affil{Department of Physics, University of Waterloo, Waterloo ON N2L 3G1}
\email{glharris@astro.uwaterloo.ca}

\begin{abstract}

The metallicity distribution function (MDF) for the old red-giant
stars in the halo of NGC 5128, the nearest giant elliptical galaxy,
is virtually identical with the MDF for the old-disk stars in the
LMC and also strongly resembles the halo MDF in M31.
These galaxies all have high mean halo metallicities  
($\langle$m/H$\rangle \simeq -0.4$) with very small proportions
of low-metallicity stars.  These observations reinforce the view that
metal-rich halos are quite normal for large galaxies of all types.  Such systems
are unlikely to have built up by accretion of pre-existing, gas-free small
satellite galaxies, unless these satellites had an extremely shallow
mass distribution ($\Delta$ log $N / \Delta$ log $M \gtrsim -1$).
We suggest that the halo of NGC 5128 is more likely to
have assembled from hierarchical merging of gas-rich lumps
in which the bulk of star formation took place during or after the merger
stage.

\end{abstract}



\section{Introduction}

Part of the early historical record of galaxy formation is left behind
in the ages and metallicities of the old stellar populations.
Much recent effort has gone into empirically
deducing the star formation history and the
age-metallicity relations (AMR) within the Local Group galaxies
\citep[see, e.g,][for an extensive review]{mat98}.
For progressively more distant systems, however, only the more luminous parts
of the color-magnitude diagram can be observed, and deriving an
accurate AMR becomes increasingly difficult.
Large ellipticals present a special challenge because none are 
close enough to the Local Group to allow direct, comprehensive probes 
into their stellar populations. 

Recently, we have used deep $(V,I)$ photometry from $HST/WFPC2$ to obtain
color-magnitude diagrams for fields 
in the halo of the nearest giant E galaxy, NGC 5128 
\citep[][hereafter Papers I and II; see also Marleau et al.~2000 for $NICMOS$ near-infrared
data to similar photometric limits]{har99,har00}.
Our two target fields are at projected distances of 21 and 31
kpc from the galaxy center and thus represent 
its remote halo, far from the more metal-rich inner
bulge and most recent star formation activity (see Paper I).
These CMDs reveal the top $\sim2.5$ magnitudes of the old-halo
red-giant branch (RGB).
The RGB stars by themselves cannot be used to
construct a true age-metallicity relation, but because the RGB loci for
old stars are far more sensitive to metallicity than age,
we can use them to construct
a first-order {\it metallicity distribution function} (MDF) 
for the NGC 5128 halo -- the first such one for any
giant elliptical and one which does not rely on indirect arguments from
luminosity-weighted integrated light.  The MDF is an important step
towards an eventual complete AMR.

Our knowledge of MDFs for the various Local Group galaxies has 
also increased dramatically in quality over the past few years,
and it is natural to ask which of these ``template'' MDFs (if any)
might resemble the halo of the giant elliptical. 
Giant E galaxies are expected to have formed through a wide range
of possible routes including hierarchical merging of gas clouds
at early times, later merging of disk galaxies, and ongoing
accretion of small satellites.  Many of these candidate
pre-galactic units (from small gas clouds to fully formed large disk
galaxies) should resemble the dwarfs and disk galaxies
found in the Local Group.  

\section{Comparisons of Metallicity Distributions}

The outer halo of NGC 5128 
has an MDF which peaks at [m/H] $\simeq -0.45$ 
and, although a low-metallicity ``tail'' is present,
the number of stars with [m/H] $< -1$ is remarkably small.
(By contrast, the halo of the Milky Way is populated almost entirely
by stars with [m/H] $< -1$.)
In Paper II, we noted that the overall shape of the NGC 5128 MDF 
is roughly similar to those of the compact dwarf 
elliptical M32 \citep{gri96}
and the halo of M31 \citep{dur01,ric01}.
Perhaps surprisingly, however, it bears an even closer resemblance
to the MDF drawn from the old stars in the Large Magellanic Cloud.

A direct comparison between NGC 5128 and the LMC
is displayed in the upper panel of Figure 1.  
For NGC 5128, we show the MDF defined by the 436 halo RGB stars 
in our database brighter than $M_I = -3.0$, i.e. the stars 
in the brightest one magnitude of the
giant branch where the RGB color range (and metallicity 
sensitivity) is largest and the spread due to photometric
measurement uncertainty is negligible (see Paper II).
For the LMC, we use the ``outer disk'' sample 
from \citet{col00a}, which includes 39 RGB stars with high-quality
spectroscopic abundances from the age-insensitive Ca II index.
MDFs for the LMC old disk derived through Str\"omgren photometric indices 
are also available for samples of many hundreds of stars
\citep[see][for comprehensive discussions]{col00a,lar00,dir00}
and these agree closely with the spectroscopically based distributions.

Two additional MDFs drawn from the Milky Way and M31
are shown in the lower panel of Figure 1.  
The halo of M31 \citep{dur01}
has an MDF with the same peak as NGC 5128
or the LMC, though with a more prominent low-metallicity tail.
The small members of
the Local Group, from the SMC downward, have lower mean abundances
by at least a factor of two, as do the halos of M33 and the
Milky Way itself \citep{rya91,pap00}.
Both the old disk and bulge of the Milky Way 
\citep{wys95,mcw94,fro99}
are two to three times more metal-rich than the NGC 5128 halo.
Finally, the compact elliptical M32, at least in the inner regions studied
by \citet{gri96}, also has a high mean metallicity
similar to the bulges of the large spirals.

Statistical comparisons of the NGC 5128 halo with M31 and the
LMC show that (a) the LMC old disk
differs from NGC 5128 at only the $\sim$45\% confidence
level (i.e., the two are indistinguishable); and (b) the M31 halo is
marginally different (95\% confidence), primarily because of its
more prominent low-metallicity tail.
It should be noted that all of these MDFs are
tied to the same Milky Way globular cluster metallicity scale through 
RGB model tracks and the fiducial loci of well studied clusters
(see the discussions of Paper II and \citeauthor{col00a}).
Formal statistical comparisons aside,
we suggest that the MDFs of the oldest stars in these three galaxies are, 
for all practical purposes, quite similar despite their 
wide variety of host galaxy type.

This combined evidence suggests that the halo of NGC 5128   
is not a relic of mergers of either large Milky-Way-like disks or bulges
(which are distinctly too metal-rich),
or dwarf stellar systems (which are too metal-poor).  By contrast, 
if many intermediate-sized
objects like the LMC were merged and their light 
allowed to age-fade for
several Gyr, the result would be very like what we now
see in the NGC 5128 halo.  In the next section we will discuss the
range of possibilities a bit further.

\section{Discussion:  Accretion or Merger?}

One view of large-galaxy formation, 
explored in detail by \citet{cot00},
is that galaxy halos built up by simple accretion
of smaller objects which had already formed 
their own stars and thus held little or no remaining
gas when they merged.  Under this assumption, the MDF of the
final halo is the straightforward result of combining
the mass distribution of the accreted satellites 
with their metallicity distribution versus mass (or luminosity).

The empirical trend of mean metallicity [m/H] with luminosity
$M_V^T$ for Local Group members is shown in Figure 2.  For all 
objects in Fig.~2, the metallicities are averages
of large samples of individual stars.
The data are taken from \citet{mat98},
with various more recent measurements for Cetus
\citep{whi99}, M32 \citep{gri96},
the Andromeda dwarf spheroidals \citep{dac00,cot99,arm98,arm99,gre99},
the SMC \citep{lar00},  Fornax \citep{sav00}, Sagittarius
\citep{bel99,lay00}, and M33 \citep{pap00}.
For the three large spirals, the metallicity
refers to the halo and, for the SMC and LMC, it refers to the
outer old-disk population.  The galaxy
luminosities are taken from the Local Group catalog of 
\citet{van00}.  There is significant scatter, but the overall
trend displayed by the dwarfs 
plausibly follows the scaling relation 
$Z \sim L^{0.37}$ predicted by \citet{dek86}.  Figure 2 
reinforces the view that for large galaxies of all types,
moderately metal-rich halos may well be the norm, not the exception. 

To explore the accretion model a bit more quantitatively, we employ a modified  
version of the numerical method of \citet{cot00}.
The relative numbers of accreted satellites are assumed to
follow a power-law distribution $dN/dL \sim L^{\alpha}$.  The  
lower limit $L_{min}$ of the distribution is set at 
$M_V^T = -8.5$, similar to the smallest dwarf
spheroidals (Draco, Ursa Minor), and the upper limit $L_{max}$ at 
$M_V^T = -19.5$, which is equivalent to about $2 \times 10^{10} M_{\odot}$
assuming $(M/L)_V=4$, or about four times the
mass of the LMC \citep[which is $\simeq 5\times 10^9 M_{\odot}$;][]{alv00}.
The upper mass cutoff by definition sets the maximum metallicity that
can be contributed by
the accreted halo stars.  Its value is a somewhat arbitrary choice,
but in setting it roughly at this 
point we are guided by recent simulations of the mass spectra
of pregalactic clouds in standard cold-dark-matter cosmologies, 
which typically give
$\alpha \simeq -1.8 \pm 0.2$ and upper ends $\sim 10^{10} M_{\odot}$
\citep[e.g.,][]{fre96,wei01}.

Next, the MDF within each accreted dwarf
is assumed to have a Gaussian form\footnote{Ideally, we might expect
the MDF for each individual dwarf to follow the 
Simple Model (closed-box enrichment)
of chemical evolution (see C\^ot\'e et al.), since in this case each
satellite is assumed to have completed its star formation in isolation.  
However, in reality many dwarf galaxies have 
much smaller relative numbers of low-metallicity 
stars than the Simple Model predicts, and their observed MDFs
more nearly approximate a symmetric Gaussian distribution
(see Paper II for additional examples).  A notable exception
to this trend is the halo of M31, in which the MDF has the 
prominent low-metallicity `tail' that rather accurately 
mimics a closed-box model with effective yield $y_e = 0.005$ 
\citep{dur01}.}  
with dispersion $\sigma$[m/H] $= 0.35$ (see C\^ot\'e et al.~regarding
the empirical evidence for a uniform $\sigma$)
and with mean $\langle$m/H$\rangle$ given by the correlation
in Fig.~2, namely $\langle$m/H$\rangle = -3.33 - 0.148 M_V^T$.
We select galaxies in equal luminosity intervals,
in proportional numbers given by the adopted power law, and
then add them together to calculate the combined MDF.
Our simulation is not intended to match the much more extensive range of
Monte Carlo realizations explored by C\^ot\'e et al., but only to
delineate changes to the first-order features of the MDF as the 
slope parameter $\alpha$ is varied.  

In Figure 3 we show the results for model halos built from
four different input mass distributions ($\alpha = -2.0, -1.8, -1.0, 0.0$).
For the steepest mass spectra (largest negative $\alpha$ values), 
the summed MDFs are
very broad and dominated by the large numbers of metal-poor dwarfs,
though the detailed shape 
is quite sensitive to $\alpha$.  Nevertheless, for any $\alpha
\lesssim -1$ the model predicts too many low-metallicity stars to
successfully match NGC 5128.

With suitable combinations of $\alpha$ and $L_{max}$ it is possible
to obtain approximate matches to a wide variety of actual MDFs.
However, for distributions as narrow and as metal-rich as we find
in NGC 5128, the range of solutions is pushed strongly towards
both a high $L_{max}$ (LMC-like or larger) and a shallow mass
spectrum $\alpha$.
For $\alpha > -1$, the model result
becomes much less sensitive to $\alpha$
because the summed MDF is dominated by the few largest satellites.
These flat input spectra can yield MDFs with approximately the right mean
and dispersion to describe NGC 5128 and the other galaxies mentioned above.
That is, under the accretion scenario, 
NGC 5128 would likely have assembled
almost entirely from rather large satellites, similar to the LMC, SMC, or
M32 in size.

Mass distributions with $\alpha > -1$ are not out of the 
question on observational grounds, but are certainly unusual.
Field galaxies and small groups 
typically show $\alpha = -1.0$ to $-1.3$ in the dwarf regime
\citep[e.g.,][]{bin88,jer00,zab00,lov00,bla01}.
Notably, the slope for the Centaurus cluster itself, within which 
NGC 5128 is the central dominant member, is near $\alpha \sim -1.3$
\citep{jer00}, too steep for the purposes of the accretion model.
In bigger clusters of galaxies, 
$\alpha-$values in the range $-1.5 \pm 0.2$ are 
frequently quoted, though the results appear to depend on limiting 
magnitude as well as environment \citep[][among others]{fer91,dep98,sec97}.
Slopes as high as $\sim -1.7$ to $-2$ have been found   
for environments within rich clusters and for the faintest part
of the dwarf sequence \citep[e.g.,][]{smi97,tre97,tre98,phi98,dep95}.
{\it If} these present-day slopes are fair indicators of the 
mass spectra at early times, then most of them would produce 
accreted halos with intermediate metallicity and a broad internal range,
but not as high a mean as in M31 or NGC 5128.

Constraints on the choice of $(L_{max}, \alpha)$ are driven by the
dispersion in the observed MDF as well as its mean.  For example,
we could produce a model MDF with a steep $\alpha \ltsim -1.5$ and with the
same {\it mean} metallicity as in NGC 5128 or M31, 
if we also increased $L_{max}$ beyond the value adopted above.
But this combination would bring in too many stars at high metallicity
(from the very most massive accreted satellites) {\it and} too many
stars at low metallicity (from the large numbers of small 
satellites), increasing the MDF dispersion beyond acceptable bounds.
Within the context of the accretion model as given, we suggest that values
of $L_{max}$ corresponding to masses $\ltsim 5 \times 10^{9} M_{\odot}$ 
or $\gtsim 5 \times 10^{10} M_{\odot}$ would not fit 
the NGC 5128 or M31 halo MDFs for any plausible slope $\alpha$. 
That is, $L_{max}$ needs to be at least as high as the LMC itself in order
to supply enough high-metallicity stars (after which $\alpha$ can be
adjusted to match the low-metallicity end correctly); but if $L_{max}$
is {\it many} times larger than the LMC, i.e.~approaching the Milky Way bulge
in size, then too many very high-metallicity stars are brought in
and no $\alpha-$slope can compensate for it.

An alternate modelling approach is to assume that the 
pregalactic lumps still had a large fraction of gas at the time 
when they were beginning to merge.  
This route opens up much more parameter space for
formation models, including such processes as
the rates of star formation within each dwarf, the 
merger rate, or interactions through winds and feedback
\citep[e.g.,][]{col00b,kau93,fre96,sca01}.
Such models provide several possibilities for
reconciling a steep initial
mass spectrum of pregalactic fragments ($\alpha \sim -2$)
with a final MDF that has moderately high mean metallicity 
and almost no low-metallicity stars.  It is not yet clear which
of these approaches would apply best in detail to NGC 5128.
A qualitative but self-consistent picture for its formation would be
that NGC 5128 built up from a moderately flat mass spectrum of subsystems
(consistent with its local observed $\alpha-$level), and that
these had already formed some stars of their own but were still mostly
gaseous.  The major stage of star formation then took place within
the much larger potential well of the new giant elliptical, which
could hold on to most of the gas and complete the conversion into
stars.  As we discuss in Papers I and II, this basic scenario has the
extra advantage that it is 
also consistent with the characteristics of the globular cluster
system in NGC 5128.

The halo of the Milky Way itself (along with the less luminous spiral M33)
is strikingly different from the other large galaxies in Fig.~2.
C\^ot\'e et al. show in detail that the
very low mean metallicity $\langle$m/H$\rangle = -1.6$
which characterizes both these spirals 
might plausibly have built up from accretion of stellar
subsystems obeying an input mass distribution 
$\alpha \sim -1.8$ to $-2.0$.  However, the observed slope
of the luminosity function for the Local Group dwarfs
is $\alpha \simeq -1.1$ \citep{pri99}, very
much at odds with this requirement.\footnote{A slope this shallow
would, however, be much closer to matching the metal-rich halo
of M31. See \citet{iba01} for new evidence 
that M31 may have tidally stripped large numbers of
stars from M32 and NGC 205, which are
moderately metal-rich.  The dichotomy between the Milky Way and
M31 halos is one of the strongest anomalies in the Local Group.}
The recent discovery of the Sagittarius dwarf now being tidally disrupted 
within the Galactic halo has been seized on as the ``smoking gun'' 
evidence that much or most of the Milky Way halo may have been accreted, 
but clearly it is an atypical event.
At $\langle$m/H$\rangle = -1.2$ and $M_V^T = -13.8$ 
\citep{bel99,lay00}, Sagittarius is considerably 
more massive and metal-rich than the average satellites could have been.

More generally, {\it if} the halos of M31 and the Milky Way 
were built by stellar accretion, this would have required strong
and apparently arbitrary differences in the type of satellite
accreted.  For M33, less can be said in detail since the 
available data are still preliminary \citep{pap00}.  However, it
is intriguing that although M33 resembles the LMC in total
size and the general age distribution of its stellar population,
its sparse and metal-poor halo appears to more
closely resemble that of the much bigger Milky Way.

The Milky Way's extremely metal-poor halo is,
in a sense, the least likely result for a big galaxy 
assembled by hierarchical merging.  If it was built from 
pure accretion, it would have required the extreme
combination of a steep input mass spectrum and
fragments which were completely stellar.  If instead it
was built from hierarchical
merging of gaseous fragments, it would
have required gas clouds which merged
into the Galactic halo but then left
almost all their original gas unused, allowing the majority 
of their gas to dissipate and
fall inward to the bulge before it could form stars 
\citep[e.g.,][]{iba95,sam97,chi01}.

The discrepancy between the theoretically expected {\it mass} spectrum
slope ($\alpha \sim -1.8$) and the much shallower values in
the {\it luminosity} distribution now observed
in environments like Centaurus and the Local Group ($\alpha \sim -1.2$)
is also a concern.  The present-day observed spectrum may, of course, not be
as steep as the original pregalactic one, perhaps because the stellar
content of the smallest ones dissipated entirely, leaving them as
only dark halos \citep[e.g.,][]{kly99}.
Thus a number of interpretive problems remain for future study.

\section{Summary}

Metallicity distribution functions based on large samples of star-by-star 
measurements are now available for several large galaxies,
as well as many dwarfs in the Local Group and elsewhere.  This material
shows that the MDFs for the oldest stars in NGC 5128, M31, and the LMC 
are metal-rich and closely similar
to one another despite their very different
environments.  With the aid of
simple numerical simulations, we argue that the halo
populations in these big galaxies are unlikely to have formed exclusively by
direct accretion of smaller stellar systems, {\it unless} these
small satellites had a unrealistically flat input mass distribution
$\alpha > -1$ or (in an even more extreme view) were all LMC-like
in size.

Hierarchical merging of gas-rich fragments would allow a much
wider range of possibilities for MDFs in these large
galaxies and make it easier to produce moderately
metal-rich systems.  
The fact that the MDFs in the large galaxies discussed
above are so similar to one another
leads us to speculate that their early histories may have 
resembled one another in ways not yet fully understood.

\acknowledgments

This work was supported by the Natural Sciences and Engineering
Research Council of Canada through research grants to the authors.


\clearpage

\figcaption[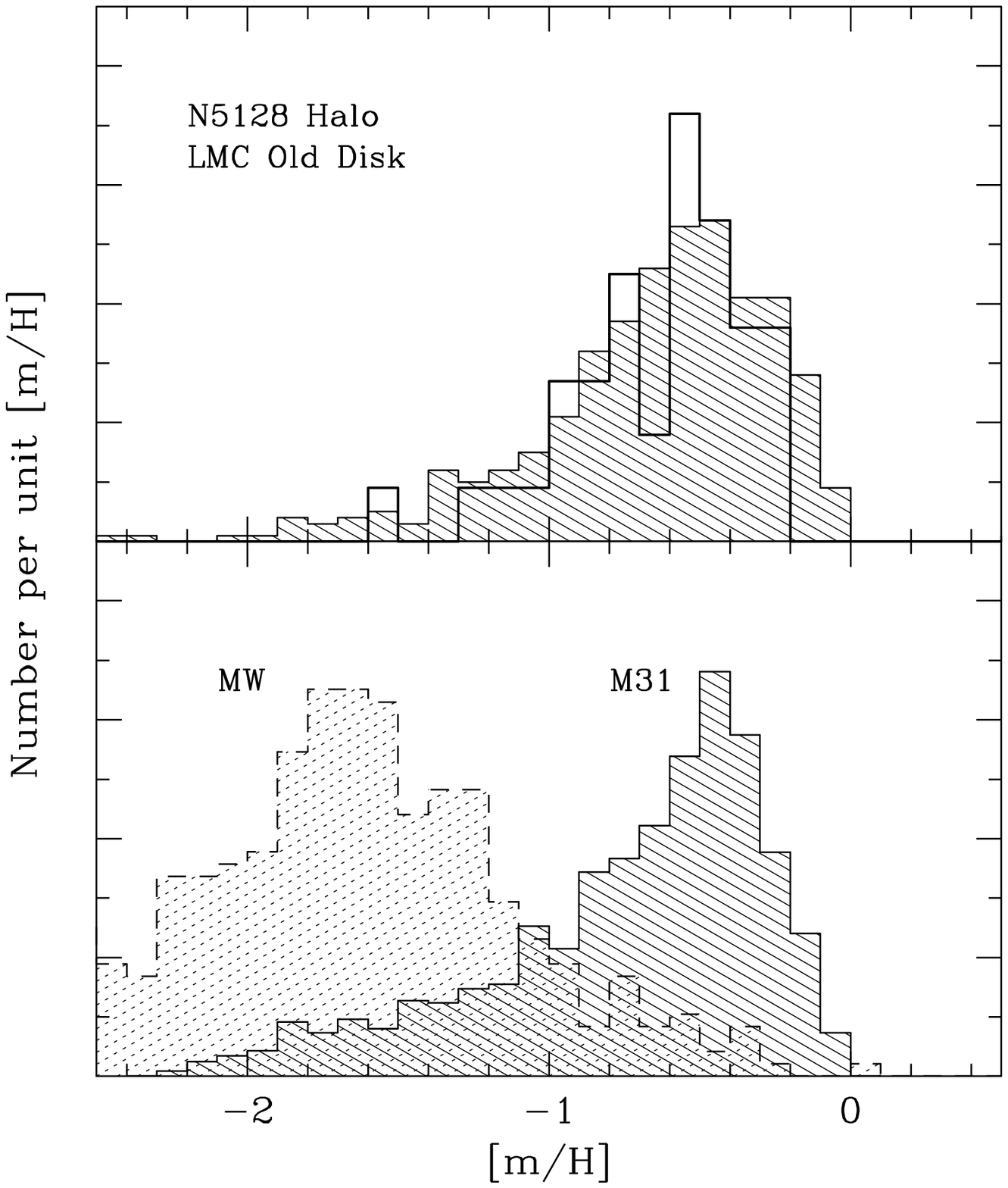]{{\it Upper panel:}  Metallicity distribution
functions for the stars in the outer halo of the giant elliptical NGC
5128 (shaded histogram) and red-giant stars in the outer disk of the
Large Magellanic Cloud (heavy solid line).  The two distributions are
statistically identical.  {\it Lower panel:}  Metallicity distributions
for stars in 
the halo of M31 (shaded histogram at right), and the halo of the
Milky Way (stippled histogram at left).  \label{fig1}}

\figcaption[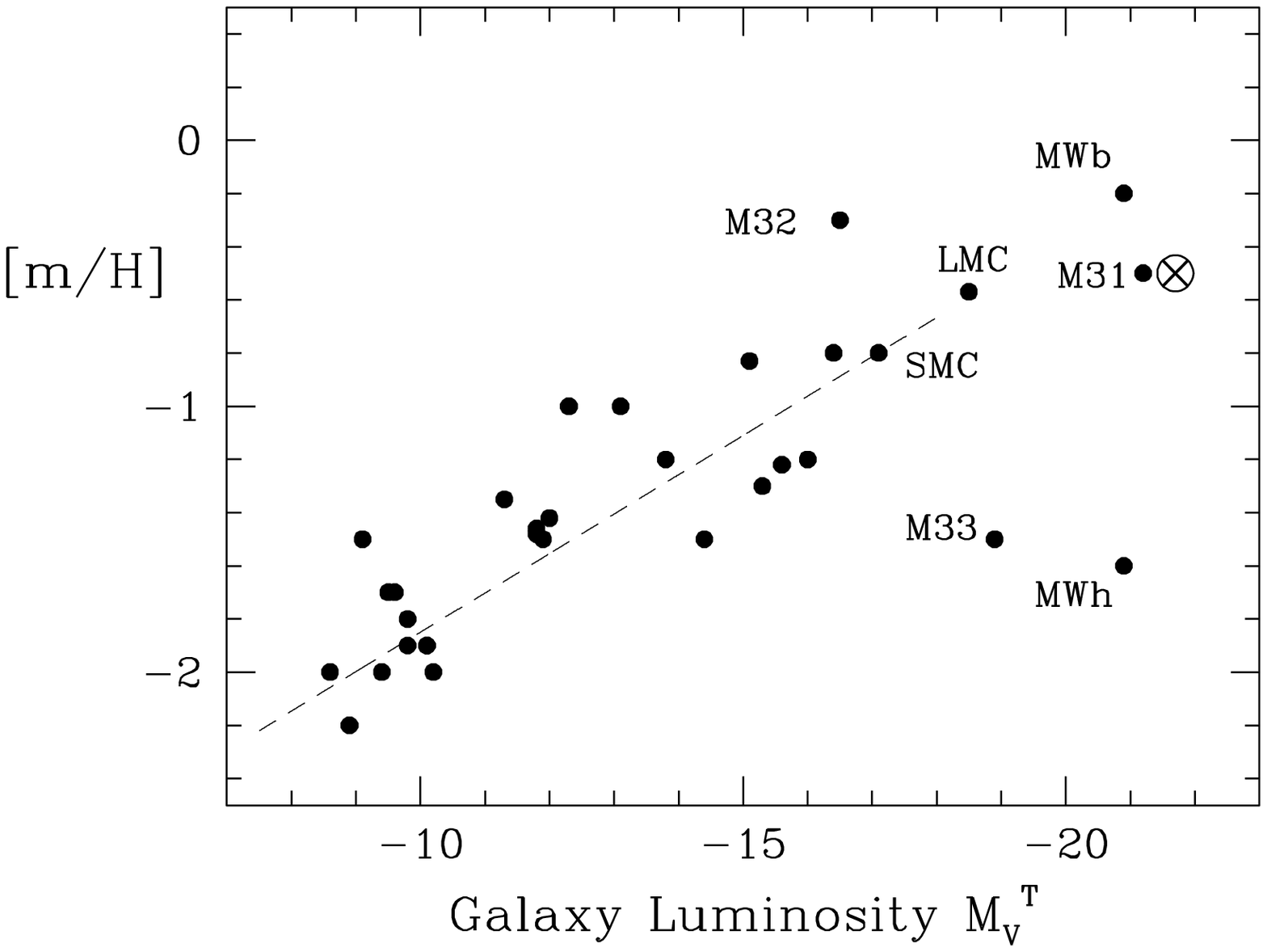]{Mean metallicity of stars in Local Group
galaxies ({\it solid dots}) plotted against 
galaxy luminosity $M_V^T$.  The {\it circled cross} denotes the 
halo of NGC 5128.  For NGC 5128,
M31, LMC, SMC, and M33 the points refer to the old-halo stars only.
The Milky Way is plotted twice, separately for its halo (MWh)
and bulge (MWb).
The expected scaling relation for dwarf galaxies formed in CDM potential
wells, $Z \sim L^{0.37}$ (Dekel \& Silk 1986; C\^ot\'e et al.~2000),
is shown as the dashed line.  \label{fig2}}

\figcaption[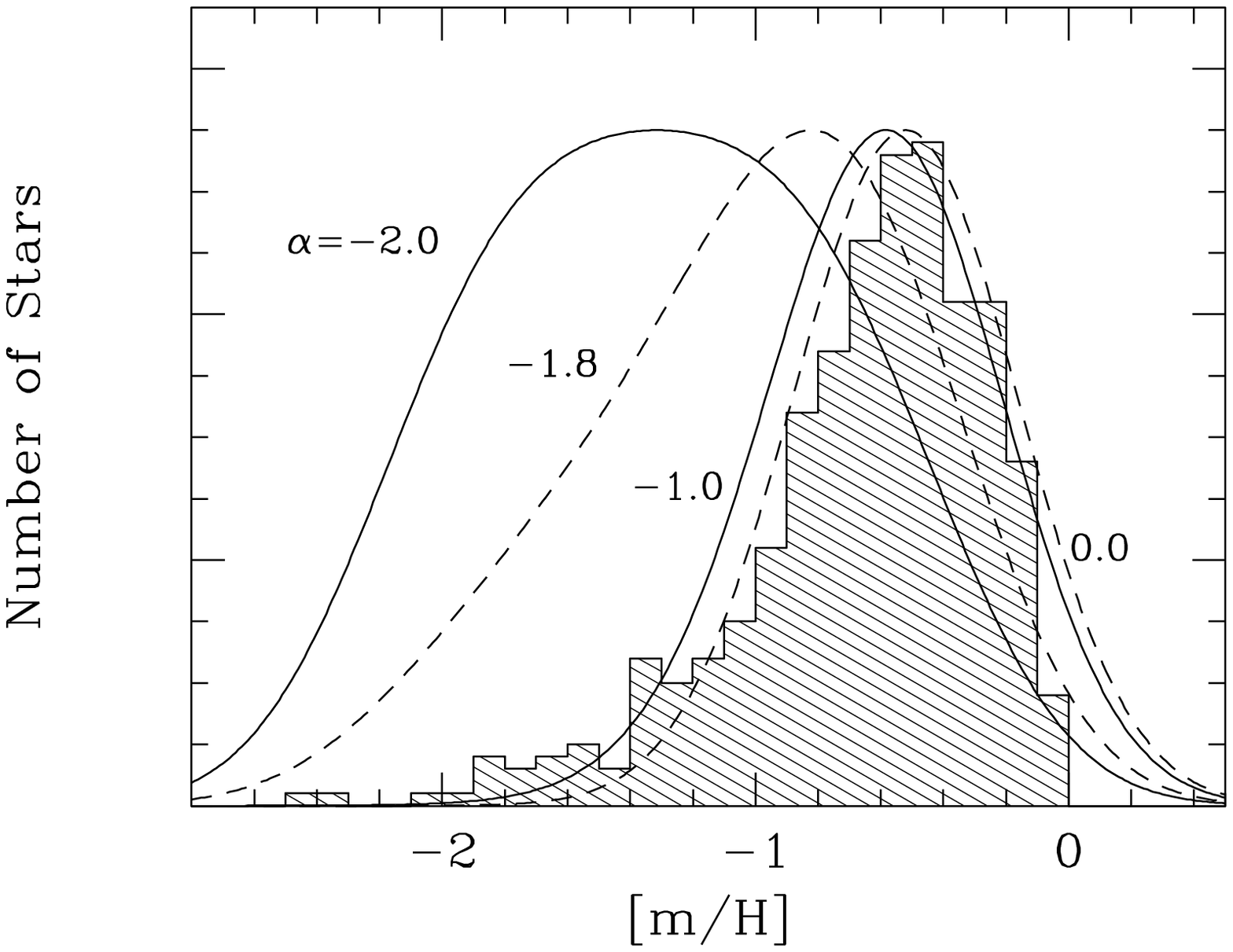]{Metallicity distribution function for
the NGC 5128 halo (shaded histogram) and for various simulated MDF
models as described in the text.  The halo is assumed to be built
up from the sum of accreted satellites following a mass 
distribution function $dN/dM \sim M^{\alpha}$.  Four illustrative models
show the average MDFs for $\alpha = -2.0$ (leftmost solid line),
$-1.8$ (middle dashed line), $-1.0$ (rightmost solid line), and
0.0 (rightmost dashed line).   \label{fig3}}

\clearpage

\end{document}